\pdfoutput=1
\documentclass[reprint,amsmath,amssymb,aps,superscriptaddress,nofootinbib,longbibliography]{revtex4-1}

\usepackage{graphicx}
\usepackage{dcolumn}
\usepackage{natmove}
\usepackage{hyperref}
\hypersetup{colorlinks,allcolors=blue}
\usepackage[capitalize]{cleveref}
\usepackage{newtxtext}
\usepackage[varvw]{newtxmath}

\input{latextuning.rty}

\begin{document}

\title{Auger recombination in Dirac materials: A tangle of many-body effects}

\author{Georgy Alymov}
\email{alymov@phystech.edu}
\affiliation{Moscow Institute of Physics and Technology, Dolgoprudny 141700, Russia}

\author{Vladimir Vyurkov}
\affiliation{Institute of Physics and Technology RAS, Moscow 117218, Russia}

\author{Victor Ryzhii}
\affiliation{Research Institute of Electrical Communication, Tohoku University, Sendai 980-8577, Japan}

\author{Akira Satou}
\affiliation{Research Institute of Electrical Communication, Tohoku University, Sendai 980-8577, Japan}

\author{Dmitry Svintsov}
\affiliation{Moscow Institute of Physics and Technology, Dolgoprudny 141700, Russia}
             
\date{Published in Phys.~Rev.~B on May 8, 2018: \href{https://doi.org/10.1103/PhysRevB.97.205411}{https://doi.org/10.1103/PhysRevB.97.205411}}             
             
\begin{abstract}
The peculiar electron dispersion in Dirac materials makes lowest-order Auger processes prohibited or marginally prohibited by energy and momentum conservation laws. Thus, Auger recombination (AR) in these materials is very sensitive to many-body effects. We incorporate them at the level of the $GW$ approximation into the nonequilibrium Green's functions approach to AR and study the role of dynamic screening, spectrum broadening and renormalization in the case of weakly pumped undoped graphene. We find that incorrect treatment of many-body effects can lead to an order-of-magnitude error in the recombination rate. We show that the AR time depends weakly (sublinearly) on the background dielectric constant, which limits the possibility to control recombination by the choice of substrate. However, the AR time can be considerably prolonged by placing graphene under a metal gate or by introducing a bandgap. With carrier cooling taken into account, our results comply with experiments on photoexcited graphene.
\end{abstract}

\maketitle
\section{\label{sec:introduction}Introduction}

Dirac materials, such as graphene~\cite{novoselov-graphene}, topological insulators~\cite{topinsulatorsreview} (e.g., CdHgTe quantum wells~\cite{Bernevig-HgTe}), and others~\cite{Dirac_materials-review}, possess an unusual low-energy electron dispersion similar to that of relativistic electrons in vacuum, thus offering an opportunity to observe relativistic quantum effects on a tabletop. Chiral anomaly~\cite{ChiralAnomaly}, Klein tunneling~\cite{Klein-graphene}, and the formation of super-critical atoms~\cite{AtomicCollapse-graphene} are among the brightest examples of high-energy physics phenomena at the nanoscale. Conversely, the effects prohibited for relativistic electrons by conservation laws are expected to be prohibited for electrons in Dirac materials. In particular, high-energy electrons in vacuum cannot generate electron-hole pairs. In other words, impact ionization of the Dirac sea is impossible~\cite{vasko-raichev}, together with its inverse, Auger recombination (AR).

A strong suppression of AR in Dirac materials suggests the possibility of long-lived population inversion and efficient far-infrared lasing~\cite{RyzhiiJAPnegative_condutivity,RyzhiiJAPInjectionLaser}. The latter facts were indeed confirmed in CdHgTe-based quantum wells, where stimulated light emission was observed up to $19.5$ $\mu$m wavelength~\cite{Morozov_APL_LWemission}, while the measured recombination lifetime reached 65 ps at a band gap of 14 meV and liquid-helium temperature~\cite{Morozov2012lifetimeCdHgTe}. Despite the energy-momentum restriction for Auger processes being lifted in the massless case, population inversion~\cite{Gierz2014} and stimulated emission~\cite{Li-pump-probe} persisting over ${\sim} 150$~fs at ${\sim} 2500$ K have also been observed in graphene.

Though the experimentally measured short nonradiative lifetime in graphene raised almost no doubt, the calculation of this time remains a challenging issue. The energy and momentum conservation laws allow AR only between carriers with collinear momenta~\cite{Foster-Aleiner} that occupy zero volume in phase space. On the other hand, massless Dirac fermions with parallel momenta have equal velocities and, therefore, interact infinitely long, which is known as the collinear scattering singularity~\cite{Fritz_PRB}. In addition, Coulomb interaction between carriers moving in parallel undergoes infinitely strong screening~\cite{Tomadin-theory} due to the collinear divergence in the Lindhard polarization of massless Dirac fermions~\cite{Finite-temperature_polarizability}. In view of the above, the calculation of the AR rate in the massless Dirac case would lead to an expression of the form $0 \times \infty / \infty^2$ if higher-order many-body effects are neglected.

These many-body effects affect AR in Dirac materials in two radically opposite ways. Carrier-carrier scattering results in a finite quasiparticle lifetime, thus softening the energy conservation restriction~\cite{Tomadin-theory}. The Coulomb renormalization of the Dirac dispersion enhances the quasiparticle velocity in the massless case~\cite{AbrikosovBeneslavski} and increases the gap in the massive case~\cite{Kotov-review}, thus further prohibiting AR~\cite{Golub-valley}. Therefore, an accurate description of AR in Dirac materials essentially requires the inclusion of carrier interaction effects.

An early theoretical study of Auger processes in gapless graphene was carried out in~\Cite{Rana-Auger}, where static screening was assumed and the indeterminate form $0 \times \infty$ was evaluated without a detailed explanation. Later, Auger processes within the same static screening approximation were included in the semiconductor Bloch equations to study carrier kinetics in graphene~\cite{Malic-kinetics,Malic-Auger} (see also \Cite{Malic2017} and references therein).

A regularization of the $0\times \infty$ form involving weak energy nonconservation (attributed to the finite quasiparticle lifetime), which was performed in~\Cite{Tomadin-theory}, justified the result of \Cite{Rana-Auger}. At the same time, it was found that the carrier kinetics within the static screening model disagrees with experiments~\cite{Tomadin-exp}, and this discrepancy was fixed by introducing dynamic screening (evaluated for slightly noncollinear processes to avoid infinite screening and get a nonzero AR rate) into the regularized expressions~\cite{Tomadin-theory}.

Introduction of dynamic screening \emph{after} the regularization was also done in \Cite{Malic_dynamic}, where the problem of infinite screening was avoided by broadening the polarizabilities. \CCite{Malic_dynamic} also provides a consistent method to calculate this broadening, in contrast to \Cite{Tomadin-theory}, where a fixed ``noncollinearity'' parameter was used. However, even this approach cannot be considered fully satisfactory, as the regularization of expressions for the AR rate should involve dynamic screening from the very beginning. Indeed, since the screening shoots up for nearly collinear processes, the use of its collinear value (i.e., the maximal one) instead of a proper average is not justified. A more important issue is that considering only collinear Auger processes, one excludes a whole class of noncollinear plasmon-assisted processes brought about by dynamic screening. Though in principle one could include them as a separate recombination channel~\cite{Rana-plasmons}, at realistic spectrum broadening plasmon-assisted and collinear processes are not well-separated and are better to be treated within a unified approach.

Surprisingly, there have been no attempts to include the renormalization of the quasiparticle spectrum in the AR rate calculations.

In this paper, we develop a method for calculation of the AR rate in Dirac materials, where AR is forbidden or almost forbidden by conservation laws. The core of our approach is the nonequilibrium Green's functions (NEGF) formalism~\cite{NEGFhandbook}, which is a standard tool to treat many-body effects in quantum kinetics~\cite{Haug} and has been successfully used to investigate AR in parabolic-band semiconductors~\cite{Ziep-Mocker,Yevick-GW_Auger,Auger_scattering}. We find that, within the $GW$ approximation~\cite{NEGF-GW}, the AR rate can be expressed as a product of intra- and interband polarizabilities timed by the square of the screened interaction and integrated over the frequencies and wavevectors of virtual photons. The information about carrier interactions is neatly absorbed into dressed polarizabilities, which appears to be more convenient than perturbative expansions of the recombination rate~\cite{Takeshima1982PRB}.

Though the developed method is general, in this paper we focus on graphene, the most studied Dirac material. We calculate the AR time in undoped graphene and study the role of screening by dielectrics and metal gates, as well as of the possible bandgap. Many-body phenomena, such as spectrum broadening/renormalization and plasmon-assisted recombination, are all shown to have a significant impact on the nonradiative lifetime. Contrary to the results obtained within the static screening model~\cite{Malic-Auger_vs_phonons}, we find that the strong intrinsic screening of collinear processes does not leave much room to control the AR rate solely by the choice of substrate. Moreover, we show that the presence of polar optical phonons in high-$\kappa$ dielectrics may even push the nonradiative lifetime in the opposite way to what might be expected. We propose another methods of controlling the AR rate in graphene and show that it can be reduced almost by an order of magnitude under a metal gate and by several orders of magnitude in gapped graphene. Our results agree with experimental studies of carrier equilibration in photoexcited graphene~\cite{Gierz2013,Gierz2014,Gierz2015,Gierz2016} if carrier cooling is properly taken into account.
\section{\label{sec:methods}Methods}

In this section, we describe the calculation of the AR rate in weakly pumped undoped graphene. We consider the stationary case, when slightly different chemical potentials for electrons and holes are maintained by a continuous pump.

\subsection{\label{sec:model}Model}

We model graphene as a system of two-dimensional Dirac fermions with the bare single-particle Hamiltonian~\cite{Kotov-review} 
\begin{eq}{hamiltonian}
    \hat{h} = v_0 (\sigma_x \hat{p}_x + \sigma_y \hat{p}_y) + m \sigma_z,
\end{eq}
where $v_0$ is the bare Fermi velocity, $\hat{\vec{p}}$ is the electron momentum, and $\sigma_{x,y,z}$ are the Pauli matrices. The last term of the Hamiltonian takes into account the possibility that a bandgap $2m$ may appear in graphene on certain substrates~\cite{Substrate-induced_gap} or upon chemical doping~\cite{Doping-induced_gap}.

Hamiltonian~\eqref{hamiltonian} yields the following energies and eigenspinors:
\begin{eq}{dispersion}
    \epsilon_{s\vec{p}} = s \sqrt{p^2 + m^2},
\end{eq}
\vspace*{-10pt}
\begin{eq}{spinors}
     \Psi_{s\vec{p}} = \frac{1}{\sqrt{2 \epsilon_{s\vec{p}} }}
    \begin{pmatrix}
    \sqrt{\epsilon_{s\vec{p}} + m}\ e^{-i\varphi/2} \\[0.5 em]
    \sqrt{\epsilon_{s\vec{p}} - m}\ e^{+i\varphi/2}
    \end{pmatrix},
\end{eq}
with $s = +1$ ($-1$) for the conduction (valence) band and $\varphi$ the angle between $\vec{p}$ and $x$-axis ($\hbar$ and $v_0$ are generally omitted throughout this article). Each of these eigenstates comes in $g=4$ independent ``flavors'', differing in valley and spin projection. For the following, we will also need the overlap factors:
\begin{eq}{overlaps}
    u^{s s'}_{\vec{p}, \vec{p}'} &=
    \lvert \Psi^{\dagger}_{s\vec{p}} \Psi^{}_{s'\vec{p}'} \rvert ^2 =
    \frac{1}{2} \left( 1 + \frac{\vec{p} \cdot \vec{p}' + m^2}{\epsilon_{s\vec{p}} \epsilon_{s'\vec{p}'}} \right).
\end{eq}

The full second-quantized Hamiltonian of Dirac fermions interacting via the instantaneous Coulomb interaction reads 
\begin{eq}{Hamiltonian}
    \hat{\cal H} &= \hat{{\cal H}_0} + \hat{\cal V}_{e-e} = \sum_{\nu,s,\vec{p}} \epsilon_{s\vec{p}} \hat{c}^{\dagger}_{\nu s\vec{p}} \hat{c}_{\nu s\vec{p}}\\
    &+\frac{1}{2} \sum_{\substack{\nu,\nu'\\ s_{1,2,3,4}\\ \vec{p},\vec{p}',\vec{q}} } V_{1234} \hat{c}^{\dagger}_{\nu s_3\vec{p}-\vec{q}} \hat{c}^{\dagger}_{\nu' s_4\vec{p}'+\vec{q}} \hat{c}_{\nu' s_2\vec{p}'} \hat{c}_{\nu s_1\vec{p}},\\
   & V_{1234} = V^R(q) \Psi^{\dagger}_{s_3\vec{p}-\vec{q}} \Psi_{s_1\vec{p}} \Psi^{\dagger}_{s_4\vec{p}'+\vec{q}} \Psi_{s_2\vec{p}'}.
\end{eq}
Here, $\hat{c}^{\dagger}$ ($\hat{c}$) are the electron creation (annihilation) operators; $\nu,\nu' = 1,2,3,4$ label fermion flavors, and $V^{R}(q)$ is the Fourier component of the Coulomb interaction potential, which reads
\begin{eq}{Coulomb}
V^{R}(q) = \frac{2\pi e^2 }{\kappa q}[ 1 - e^{-2qd_g}]
\end{eq}
for a two-dimensional graphene sheet embedded in a material with dielectric constant $\kappa$ and placed under a metal gate at distance $d_g$.

We note that the interacting electrons can belong to different valleys, i.e., $\nu$ and $\nu'$ are allowed to be different in \eqref{Hamiltonian}. This process being perfectly legitimate was overlooked in some of the previous works~\cite{Rana-Auger,Tomadin-theory}, resulting in a missing factor of two in the expressions for the AR rate.

When graphene is placed on or in a polar dielectric, an important additional recombination channel may be provided by the polar phonons in the dielectric~\cite{Rana-plasmons}. In principle, they could be treated by adding two more terms into the Hamiltonian---namely, the bare phonon term and the Fr\"ohlich electron-phonon interaction~\cite{Jena_SPP}. However, in the Green's function language, phonons just add a new (phonon) propagator to the usual Coulomb interaction, and the interaction becomes additionally screened by lattice vibrations, $V^{R}(q,\omega) = 2\pi e^2 /[\kappa(\omega) q]$, with frequency-dependent dielectric function $\kappa(\omega)$ given by the Lorentz oscillator model (see Appendix~\ref{sec:dielectric}).

Other mechanisms of recombination typically proceed on much longer timescales and are not considered in this paper, though it is straightforward to include them if necessary. In particular, recombination through intrinsic phonons has been shown to take tens of picoseconds at room temperature~\cite{Rana-phonons}, while the radiative recombination time can be estimated as~\cite{Vasko-Ryzhii-radiative} 
\begin{eq}{radiative}
\tau_{\text{rad}}\sim\left(\frac{e^2 \sqrt{\kappa}}{\hbar c}\right)^{-1}\frac{c^2}{v^2_0} \frac{\hbar}{kT}
\end{eq}
and is around hundreds of nanoseconds at room temperature.
\subsection{\label{sec:NEGF}NEGF approach to Auger recombination}

\subsubsection{\label{sec:GW}Self-consistent \texorpdfstring{$GW$}{GW} scheme}

Within the nonequilibrium Green's functions formalism~\cite{NEGFhandbook}, the central quantities are four kinds of Green's functions: retarded/advanced, $G^{R/A}_s(p,E)$, and lesser/greater, $G^{</>}_s(p,E)$. The first two are related to each other, $G^A_s(p,E) = (G^R_s(p,E))^*$, and to the spectral function $A_s(p,E) = -(1/\pi) \Im {G^R_s(p,E)}$, thus providing information about the quasiparticle spectrum. The last two describe the distribution of occupied and empty states across this spectrum and can be expressed in terms of the distribution function ${\cal F}_s(p,E)$:
\begin{eq}{G<>}
     G^{<}_s(p,E) &= 2\pi i {\cal F}_s(p,E) A_s(p,E), \\
    G^{>}_s(p,E) &= 2\pi i ({\cal F}_s(p,E)-1) A_s(p,E).
\end{eq}
In particular, the electron/hole densities are
\begin{eq}{n}
     n_e = -ig \sum_{\vec{p},E} G^{<}_c(p,E),\quad  n_h = ig \sum_{\vec{p},E} G^{>}_v(p,E),
\end{eq}
where $\sum\limits_{\vec{p},E}{} \equiv \int {\frac{d^2 \vec{p}}{(2 \pi)^2} \frac{d E}{2 \pi}}{}$.

In the equilibrium noninteracting case, the Green's functions read
\begin{eq}{G0}
     &{G^{0\, R/A}_s(p,E) = \frac{\Theta(\Lambda - p)}{E - \epsilon_{s\vec{p}} \pm i0}}, \\
    &{G^{0\, <}_s(p,E) = 2\pi i f(E) \delta(E - \epsilon_{s\vec{p}}) \Theta(\Lambda - p)}, \\
    &{G^{0\, >}_s(p,E) = 2\pi i (f(E)-1) \delta(E - \epsilon_{s\vec{p}}) \Theta(\Lambda - p)}
\end{eq}
with $f(E) = \left[\exp(E/kT) + 1\right]^{-1}$ (remember that we consider undoped graphene).
Here, we have introduced a sharp momentum cutoff $\Lambda$ necessary to obtain a finite self-energy. Physically, it reflects the finite extent of the Dirac cones and by the order of magnitude coincides with the size of the first Brillouin zone~\cite{Kotov-review}. We have chosen $v_0 = 0.85 \times 10^6$~m/s and $\Lambda = 2.5$~eV, following \Cite{Elias}.

Many-body interactions are absorbed into the retarded/advanced self-energies $\Sigma^{R/A}_s(p,E)$, which enter the interacting (``dressed'') Green's functions:
\begin{eq}{GRA}
    G^{R/A}_s(p,E) = \frac{\Theta(\Lambda - p)}{E - \epsilon_{s\vec{p}} - \Sigma^{R/A}_s(p,E) \pm i0}.
\end{eq}
Their real and imaginary parts describe the renormalization and broadening of the quasiparticle spectrum, while their energy dependence may lead to additional spectral features like plasmon satellites observed in doped graphene~\cite{Bostwick}. $G^{</>}_s(p,E)$ can be calculated through \eqref{G<>} with the same equilibrium distribution function ${\cal F}_s(p,E) = f(E)$.

Now, the full complexity of the many-body problem is encoded in the self-energies, and we need to resort to approximations in order to make the self-energy calculations feasible. We have chosen the advanced yet tractable self-consistent $GW$ approximation, which captures all the many-body effects we are going to consider.

\begin{fig}{GW}{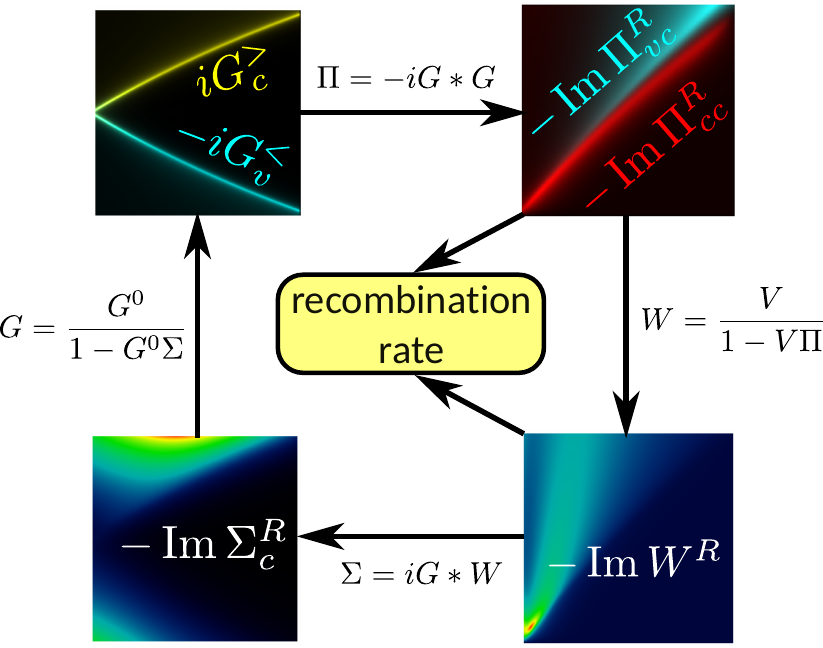}
Schematic of the present approach. First, $GW$ equations are solved iteratively to calculate inter-/intraband polarizabilities and screened interaction with many-body corrections, which are then used to find the recombination rate. For illustrative purposes, some quantities are shown on the momentum-energy plane: lesser/greater Green's functions $G^{</>}$ show the distribution of filled/empty states; imaginary parts of inter-/intraband polarizabilities $\Im \Pi^{R}_{vc/cc}$ indicate regions of inter-/intraband excitations; imaginary part of the screened interaction $\Im W^{R}$ provides information about collective excitations; and imaginary part of the self-energy $\Im \Sigma^{R}$ is minus the scattering rate.
\end{fig}

Calculations within this approximation proceed as follows~\cite{NEGF-GW} (see \cref{fig:GW}). The retarded self-energy is approximated as
\begin{eq}{SigmaGW}
    \Sigma^R_s(\vec{p},E) = i \sum_{s',\vec{q},\omega} \left( {G^R_{s'}(\vec{p}+\vec{q},E+\omega) W^{<}(\vec{q},\omega) u^{s s'}_{\vec{p},\vec{p}+\vec{q}}} \right. \\
    + \left. {G^{<}_{s'}(\vec{p}+\vec{q},E+\omega) W^{A}(\vec{q},\omega) u^{s s'}_{\vec{p},\vec{p}+\vec{q}}} \right)
\end{eq}
(Hartree contributions from electrons and holes mutually cancel in neutral graphene).

The advanced screened interaction $W^{A}(\vec{q},\omega)$, as the complex conjugate of the retarded screened interaction $W^{R}(\vec{q},\omega)$, is related to the retarded polarizability $\Pi^{R}(\vec{q},\omega)$:
\begin{eq}{WR}
    W^{A}(q,\omega)^{*} = W^{R}(q,\omega) &= \frac{V^{R}(q,\omega)}{\epsilon^{R}(q,\omega)}\\
    &= \frac{V^{R}(q,\omega)}{1 - V^{R}(q,\omega)\Pi^{R}(q,\omega)},
\end{eq}
while the lesser screened interaction $W^{<}(\vec{q},\omega)$ is related to the lesser polarizability $\Pi^{<}(\vec{q},\omega)$:
\begin{eq}{W<}
    W^{<}(q,\omega) = \Pi^{<}(q,\omega) \lvert W^{R}(q,\omega) \rvert^2  + \frac{V^{<}(q,\omega)}{\left| \epsilon^R(q,\omega) \right|^2}.
\end{eq}
Here, the second term is related to the substrate polar phonons, which are assumed to have the Bose-Einstein distribution $n_B(\omega) = \left[\exp(\omega/kT) - 1\right]^{-1}$, so $V^{<}(q,\omega) = 2in_B(\omega)\Im V^{R}(q,\omega)$ (cf. \Eqref{G<>}; $-\Im V^{R}(q,\omega)$ plays the role of a spectral function for electromagnetic modes in the dielectic, including optical phonons).

The polarizabilities, in turn, are approximated as $\Pi^{R/<}(\vec{q},\omega)=\sum_{s,s'}\Pi^{R/<}_{ss'}(\vec{q},\omega)$,
\begin{eq}{PiRGW}
     \Pi^R_{ss'}(\vec{q},\omega) &= -ig \sum_{\vec{p},E} \left({G^R_{s'}(\vec{p}+\vec{q},E+\omega) G^{<}_s(\vec{p},E) u^{s s'}_{\vec{p},\vec{p}+\vec{q}}} \right. \\
    &+ \left. {G^{<}_{s'}(\vec{p}+\vec{q},E+\omega) G^A_s(\vec{p},E) u^{s s'}_{\vec{p},\vec{p}+\vec{q}}} \right),
\end{eq}
and
\begin{eq}{Pi<GW}
    \Pi^{<}_{ss'}(\vec{q},\omega) = -ig \sum_{\vec{p},E} {G^{<}_{s'}(\vec{p}+\vec{q},E+\omega) G^{>}_s(\vec{p},E) u^{s s'}_{\vec{p},\vec{p}+\vec{q}}},
\end{eq}
closing the self-consistency cycle (note that conventions on signs and ordering of the band indices may differ in the literature).

Actual calculations are performed in reverse order, starting from noninteracting Green's functions and iterating \cref{eq:G<>,eq:GRA,eq:SigmaGW,eq:WR,eq:W<,eq:PiRGW,eq:Pi<GW} until convergence is achieved: $G^0 \rightarrow \Pi^0 \rightarrow W^0 \rightarrow \Sigma^0 \rightarrow G^1 \rightarrow \Pi^1 \rightarrow W^1 \rightarrow \Sigma^1 \rightarrow ...$

All the integrals can be cast in the form of convolutions and efficiently evaluated via the fast Fourier transform between frequency/time~\cite{Godby-space-time} and the fast Hankel transform between momentum/position~\cite{Talman-Hankel}.

\subsubsection{\label{sec:rate}Calculation of recombination rate}

In the case of a weak pump, the recombination rate can be found using equilibrium Green's functions calculated self-consistently as described in the previous section (the case of a strong pump is more complicated and remains beyond the scope of this article, see \hyperref[sec:conclusions]{Conclusions}). A weak pump does not alter significantly the quasiparticle spectrum, but changes the distribution of carriers to that with separate Fermi levels $\mu_c = -\mu_v \ll kT$: 
\begin{eq}{quasiequilibrium}
{\cal F}_s(p,E) = \frac{1}{\exp[(E - \mu_s)/kT] + 1}    
\end{eq}
(the use of the Fermi-Dirac distributions with band-dependent Fermi levels is justified by the fast intraband equilibration compared to recombination). That means that $G^{R/A}_s(p,E)$ remain unchanged, while $G^{</>}_s(p,E)$ are updated according to \eqref{G<>}, and the carrier density increases by $n_{\text{noneq}} \equiv n - n_{\text{eq}}$. The populations of each band $n_e(t)$, $n_h(t)$ are kept constant by the balance between the pump and recombination. Omitting pump terms from the Kadanoff-Baym (quantum kinetic) equations~\cite{Haug}, we obtain the total recombination rate~\cite{Ziep-Mocker,Yevick-GW_Auger}:
\begin{eq}{Kadanoff-Baym}
     R &= -\frac{\partial n_e(t)}{\partial t}  \\
    &= g \sum_{\vec{p},E} \left[ G^{<}_c(p,E)\Sigma^{>}_c(p,E) - G^{>}_c(p,E)\Sigma^{<}_c(p,E) \right].
\end{eq}

This expression encompasses all the types of recombination considered in the present model---namely, AR and recombination via substrate phonons. We stress that in our model, recombination via emission of plasmons~\cite{Rana-plasmons} should not be added as a separate channel, since we do not consider their coupling to free electromagnetic modes and assume all of them eventually decay into single-particle excitations.

Within the $GW$ approximation, the AR rate can also be written as~\cite{Ziep-Mocker}
\begin{eq}{ImPccImPcv}
    R_{\text{Auger}}^{GW} &= 8 \sum_{\vec{q},\omega} \left[n_B(\omega - \Delta\mu_{cv}) - n_B(\omega) \right]\\
    &\times \Im \Pi^{R}_{cc}(q,\omega) \Im \Pi^{R}_{vc}(q,\omega) \left| W^R(q,\omega) \right|^2\\
    &+ 4 \sum_{\vec{q},\omega} \left[n_B(\omega - \Delta\mu_{cv}) - n_B(\omega + \Delta\mu_{cv}) \right]\\
    &\times \Im \Pi^{R}_{cv}(q,\omega) \Im \Pi^{R}_{vc}(q,\omega) \left| W^R(q,\omega) \right|^2
\end{eq}
($\Delta\mu_{ss'} \equiv \mu_{s} - \mu_{s'}$). The first term describes the usual AR ($cc \leftrightarrow vc$ and $cv \leftrightarrow vv$ processes, which have equal rates in undoped graphene due to the electron-hole symmetry), the second one describes strongly suppressed by conservation laws ``double recombination'' ($cc \leftrightarrow vv$ process).

\EEqref{ImPccImPcv} with free-particle polarizabilities and dielectric function might be obtained from Fermi's golden rule. However, in our method \Eqref{ImPccImPcv} includes dressed polarizabilities, which conveniently absorb all interaction effects within the $GW$ approximation.
 
Using \Eqref{ImPccImPcv}, it is easy to observe the pathology of zero phase space and infinite interaction time, as well as the method for curing it. When many-body corrections are neglected, $\Im \Pi^{R}_{cc}$ is nonzero in the region of intraband excitations, $-q \leq \omega \leq q$, while $\Im \Pi^{R}_{vc}$ is nonzero in the region of interband excitations, $\omega \geq q$. These regions overlap only along the line $\omega = q$, where both polarizabilties have square-root singularities~\cite{Finite-temperature_polarizability}. Therefore, one faces the integration of an infinite function over a region of zero measure. It has been shown that this $0\times\infty$ limit can be found if one allows for infinitesimal energy non-conservation, the result being~\cite{Rana-Auger,Tomadin-theory}
\begin{eq}{regularizedrate}
    R_{\text{Auger}} &= g^2 v_0 \left(1-e^{-\frac{\Delta\mu_{cv}}{kT}}\right) \int_0^{\infty}{\frac{dp_{1234}}{(2\pi)^4}} \sqrt{p_1p_2p_3p_4} \\ 
    &\times f(p_1)f(p_2)f(p_3)f(-p_4) |W^R(p_1+p_3,p_1+p_3)|^2 \\
    &\times 2\pi\delta(p_1+p_2+p_3-p_4)
\end{eq}
if the exchange terms are neglected (they are not captured by the $GW$ approximation).

The situation is further complicated when one remembers that the Coulomb interaction is screened by the same divergent polarizabilities, so \Eqref{regularizedrate} with dynamic screening yields zero AR rate~\cite{Tomadin-theory}. Naturally, these singularities are washed out when quasiparticles acquire a finite lifetime.

Looking at \Eqref{ImPccImPcv} with dressed polarizabilities and screened interaction, one could identify two regions in the $(q,\omega)$ plane which provide the largest contributions to the AR rate. The first one is the region of collinear processes, $\omega \approx q$, where $\Im \Pi^{R}_{cc} \times \Im \Pi^{R}_{vc}$ is maximal. The second one lies around the plasmon dispersion, where $W^R$ is large, and can be called the region of plasmon-assisted recombination. In practice, when a realistic quasiparticle lifetime is used, these two regions usually become broad enough to be indistinguishable. They may become better separated in doped graphene, where the plasmon spectrum is sharper and lies farther away from the $\omega=q$ line.

In polar substrates, there are surface polar phonon modes (SPPs) that are the origin of dispersion and intrinsic dielectric loss ($\Im \kappa(\omega) > 0$) and provide an additional recombination channel~\cite{Rana-plasmons}:
\begin{eq}{R_phonon}
    R_{\text{SPP}}^{GW} &= 4 \sum_{\vec{q},\omega} \left[n_B(\omega - \Delta\mu_{cv}) - n_B(\omega) \right]\\
    &\times \Im \Pi^{R}_{vc}(q,\omega) \frac{\Im V^{R}(q,\omega)}{\left| \epsilon^R(q,\omega) \right|^2}.
\end{eq}
In practice, the separation of the total recombination rate into Auger and SPP contributions is not fully justified, because the interaction between SPPs and carriers in graphene gives rise to additional branches of collective excitations~\cite{Jena_SPP,Rana-plasmons}, which enhance AR via anti-screening ($\left| \epsilon^R(q,\omega) \right| < \left| \epsilon^R(q,0) \right|$) the same way as ordinary plasmons do.
\section{\label{sec:results}Results and discussion}

\subsection{\label{sec:mainresults}Auger recombination in graphene: Role of dielectric environment, gate, and bandgap}

\begin{fig}{dimensionless}{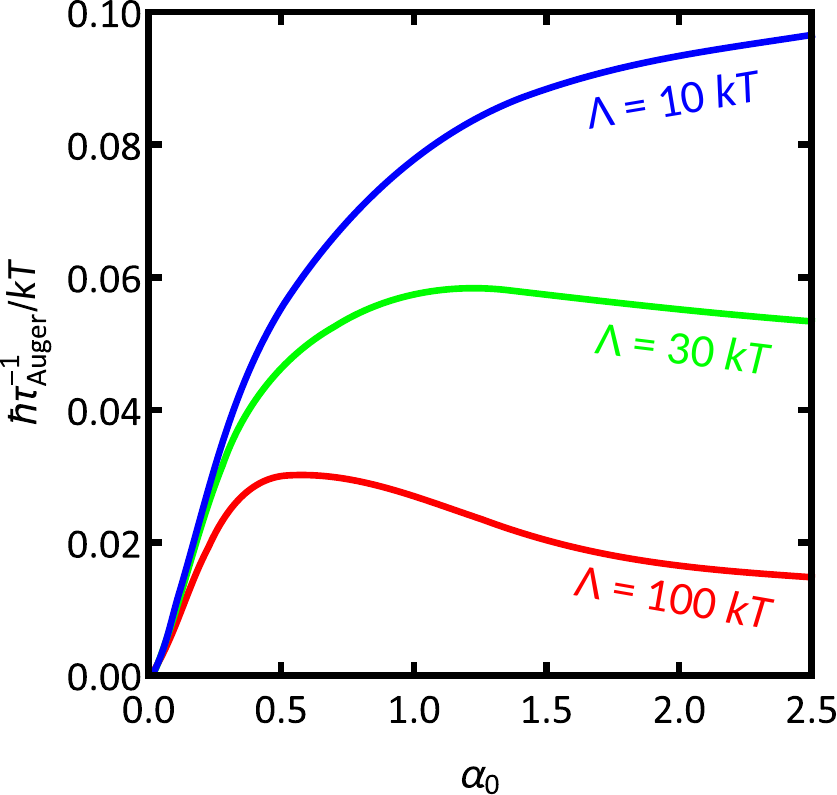}
Dimensionless Auger recombination rate in weakly pumped undoped graphene vs the bare coupling constant $\alpha_0 = e^2/(\hbar v_0 \kappa)$ at different values of the ultraviolet cutoff $\Lambda$. From now on, all the results are obtained from \Eqref{ImPccImPcv} with $GW$ polarizabilities as described in \hyperref[sec:methods]{Methods}, unless indicated otherwise.
\end{fig}

\begin{fig}[!b]{substrate}{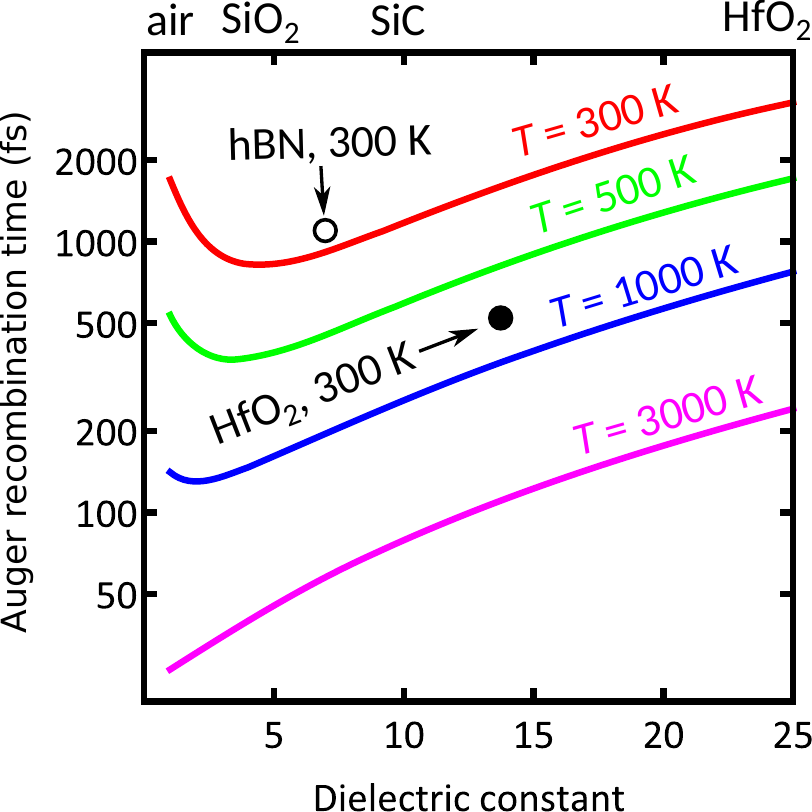}
Auger recombination time in weakly pumped undoped graphene vs the background dielectric constant at different temperatures. Static dielectric constants of several common materials are indicated on the top. Circles denote full recombination times [\Eqref{ImPccImPcv} + \Eqref{R_phonon}] calculated using frequency-dependent dielectric functions of hafnium dioxide (filled circle) and hexagonal boron nitride (open circle). These dielectric functions are presented in Appendix~\ref{sec:dielectric}. The dielectric function of HfO$_2$ is taken from \protect\Cite{HfO2kappa}, where the static value $\kappa_0 \approx 14$ was obtained instead of the more common value $\kappa_0 \approx 25$.
\end{fig}

We first present our results for the simplest case of graphene in a dispersionless medium with dielectric constant $\kappa$ at temperature $T$. Using dimensional arguments, it is possible to show that the AR time $\tau_{\text{Auger}} = n_{\text{noneq}}/R_{\text{Auger}}$, when expressed in dimensionless form $\hbar \tau^{-1}_{\text{Auger}}/kT$, can depend only on two dimensionless parameters: the bare coupling constant $\alpha_0 = e^2/(\hbar v_0 \kappa)$ and the ultraviolet cutoff expressed in units of temperature, $\Lambda/kT$ (\cref{fig:dimensionless}). One might anticipate $\tau^{-1}_{\text{Auger}} \propto \alpha^2_0$ because the AR rate is proportional to the Coulomb interaction squared. However, this is realized only at very small $\alpha_0$. At intermediate $\alpha_0$, screening by intrinsic carriers comes into play, reducing the role of substrate, and linear behavior is observed instead. At large $\alpha_0$, the AR rate can even drop down, and a pronounced dependence on the ultraviolet cutoff $\Lambda$ emerges. This feature is related to the slowly decaying tails of the quasiparticle spectral function, which introduce a cutoff dependence into the intraband polarizability and, consequently, the screened interaction (see the discussion at the end of this section).

In \cref{fig:substrate} we show the same dependence in dimensional variables,\footnote{Due to large computational demands, $T = 300$~K and $T = 500$~K results were obtained by extrapolation. The introduced error is comparable to the errors arising from uncertainty of the cutoff and the neglect of vertex corrections.} assuming $\Lambda$ = 2.5 eV~\cite{Elias}. At room temperature, the AR time is around 1--2 ps, while at elevated temperatures $T=1000$--3000~K typical for carriers in photoexcited graphene it can be as low as several tens of femtoseconds. Remarkably, the AR rate does not change with varying the background dielectric constant $\kappa$ as much as one could expect. Increasing $\kappa$ from 5 to 25 leads only to a fourfold suppresion of AR at room temperature. In practice, embedding graphene into high-$\kappa$ dielectrics can even shorten the recombination time due to the emerging channels of recombination with emission of polar substrate phonons. This shortening for graphene embedded in hafnium dioxide is illustrated in \cref{fig:substrate} by the filled circle. In contrast, optical phonons in intermediate-$\kappa$ dielectrics (e.g., hexagonal boron nitride) usually lie at higher energies and have smaller oscillator strengths, therefore having only a minor impact on AR at room temperature. This is demonstrated with an empty circle in \cref{fig:substrate}; AR is slightly suppressed in this case due to the increase of the dielectric function near the phonon energy.

\begin{fig}{gate}{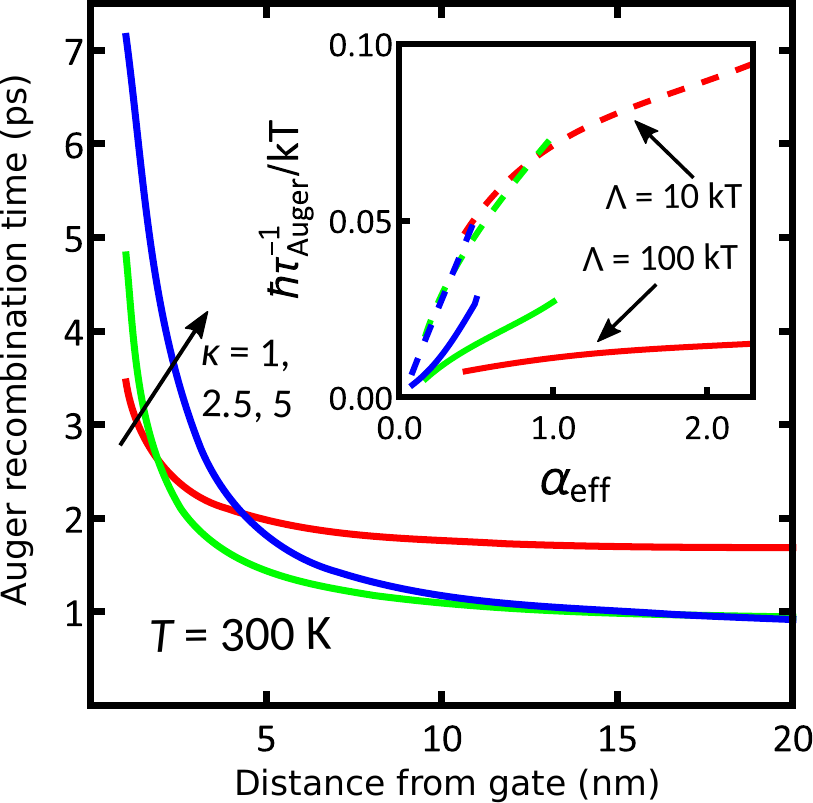}
Auger recombination time in weakly pumped undoped graphene vs the distance from a metal gate at room temperature and different dielectric constants $\kappa$. Inset: 
normalized recombination rate $\hbar \tau^{-1}_{\rm Auger}/kT$ vs effective coupling constant including the effect of gate screening $\alpha_{\text{eff}}=\alpha_0 [1-\exp(-2q_T d_g)]$, where $q_T = 2kT$ is the ``thermal wave vector''. Solid lines: $\Lambda = 2.5$~eV; dashed lines: $\Lambda = 0.25$~eV.
\end{fig}

Now we move on to the case of graphene under a metal gate, which can serve as an alternative source of screening. As shown in \cref{fig:gate}, in this way the AR rate can be reduced by almost an order of magnitude. The form of the modified Coulomb interaction $V^{R}(q) = (2\pi \alpha_0/q) \left[ 1 - \exp(-2qd_g) \right]$ suggests it might be possible to absorb the influence of the gate into an effective coupling constant $\alpha_{\text{eff}}=\alpha_0 [1 - \exp(-2q_T d_g)]$ with $q_T$ being some characteristic wavevector of order $kT$, and $d_g$ being the distance from the gate. This is indeed possible at low cutoff, as shown by the dashed lines in the inset. However, it becomes impossible at larger cutoff (solid lines in the inset), because the mechanism of the cutoff dependence of the AR rate involves also processes with large momentum transfer $q \gg q_T$, and the above arguments do not apply.

\begin{fig}{gap}{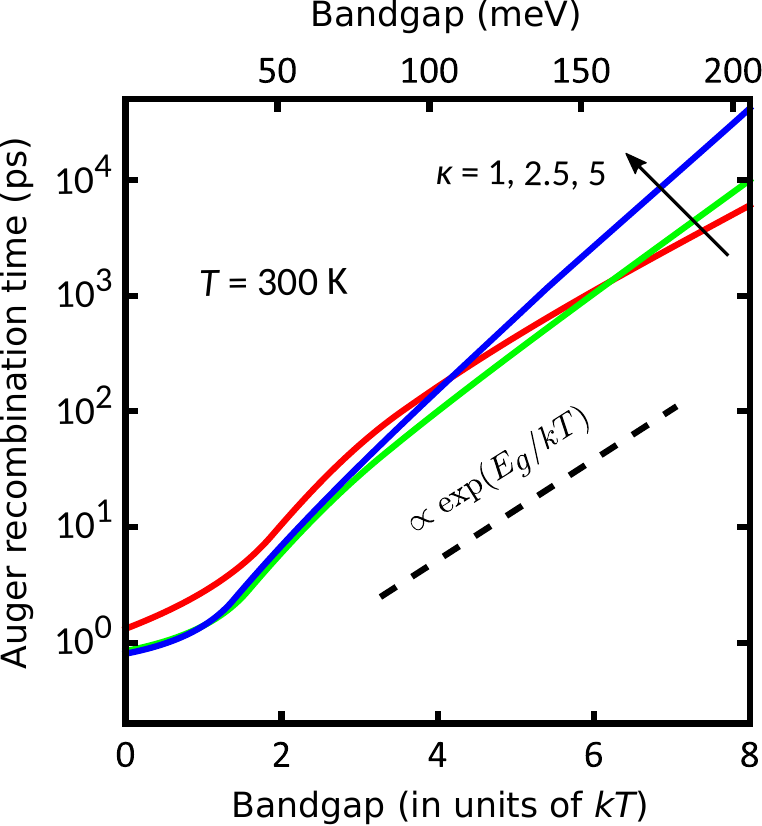}
Auger recombination time in weakly pumped undoped graphene vs the (renormalized) bandgap at room temperature and different dielectric constants $\kappa$. The dashed line shows $\tau_{\text{Auger}} \propto \exp(E_g/kT)$ law.
\end{fig}

Another option to control the AR rate in graphene is to introduce a bandgap, e.g., by substrate~\cite{Substrate-induced_gap} or chemical doping~\cite{Doping-induced_gap}. As can be seen from \cref{fig:gap}, at small gaps $E_g<kT$ the effect is minor, but at large gaps the AR time scales according to the Arrhenius law, $\tau^{-1}_{\text{Auger}} \propto \exp(-E_{\text{th}}/kT)$ (a typical behavior in gapped semiconductors: carriers involved in AR cannot have arbitrarily small energies~\cite{Abakumov-nonradiative}). The threshold energies $E_{\text{th}}$ are close to $E_g$ as a result of a competition between two factors. On one hand, conservation laws favor AR processes involving high-energy carriers, which lie on the approximately linear part of the spectrum. Near the band extrema, AR requires carriers from tails of the spectral function and is therefore power-law suppressed (assuming the spectral function is Lorentzian). On the other hand, at high energies the number of carriers drops according to the Boltzmann distribution, and AR is suppressed much more strongly (exponentially). In principle, spectrum broadening can make AR thresholdless with a non-exponential dependence on temperature and bandgap~\cite{Auger_scattering}, but at the bandgap values presented in \cref{fig:gap} the regime where carriers from deep inside the gap contribute significantly to the AR rate is not reached yet. The exponential scaling of the AR rate with the bandgap allows it to be reduced by several orders of magnitude, though gapped graphene may be unsuitable for some applications (e.g., low-THz generation).

\subsection{\label{sec:theory_vs_exp}Comparison with experiments}

Carrier equilibration in photoexcited graphene has been investigated experimentally in a number of works, including time- and angle-resolved photoemission~\cite{Gierz2013,Gierz2014,Gierz2015,Gierz2016}, pump-probe~\cite{Li-pump-probe}, and time-resolved photoluminescence~\cite{Koyama-photoluminescence} measurements. According to the series of papers by Gierz \textit{et~al.}~\cite{Gierz2013,Gierz2014,Gierz2015,Gierz2016}, who were able to track the time evolution of the carrier distrubutions in graphene after photoexcitation, the carrier dynamics in photoexcited graphene proceeds in the following stages: (i) thermalization within each band within a few tens of femtoseconds; (ii) merging of the chemical potentials for electrons/holes, typically within ${\sim} 130$ fs; (iii) cooling, which follows a biexponential decay with ${\sim} 100$ fs and ${\sim} 1$ ps time constants. The ${\sim} 130$ fs lifetime of the quasiequilibrium state with population inversion is corroborated by pump-probe experiments~\cite{Li-pump-probe}, in which negative optical conductivity persisting over a similar timescale was found. In \Cite{Koyama-photoluminescence} the authors found that carriers are not yet fully thermalized 300 fs after photoexcitation, but the carrier temperature ($T \approx 400$~K) in their experiments was significantly lower than in those mentioned above.

At typical parameters realized in these experiments (high-frequency $\kappa \approx 3.8$ for graphene on SiC, $T \approx 2000$ K), our calculations yield $\tau_{\text{Auger}} \approx 60$ fs. Inclusion of recombination via polar phonons in the SiC substrate does not significantly change this estimate. At first, it seems to contradict the experimental value of ${\sim} 130$ fs. This discrepancy is spurious, as the experimental time describes the decay of population inversion (i.e., the merging of quasi-Fermi levels), not of the nonequilibrium carrier density. This difference is important at strong nonequilibrium (when the nonequilibrium carrier density depends on the quasi-Fermi levels nonlinearly) and/or in the presence of carrier cooling.

Indeed, the density of nonequilibrium electrons is a growing function of both the quasi-Fermi level and temperature. In the absence of recombination, cooling pushes the quasi-Fermi level up to keep the carrier density constant. In the presence of recombination, the merging of electron and hole quasi-Fermi levels is slowed down by cooling. Therefore, the recombination time $\tau_r$ (which characterizes the decay of nonequilibrium carrier density) is shorter than the merging time of quasi-Fermi levels $\tau_{\Delta \mu}$ (the lifetime of population inversion).

To put this on mathematical grounds, we present a simple model that provides the means to extract the true recombination time from experimental data. We introduce four quantities: the carrier temperature $T$, the difference between the electron and hole quasi-Fermi levels $\Delta\mu$, their merging time $\tau_{\Delta\mu} = -\Delta\mu\left(d\Delta\mu/dt\right)^{-1}$, and the instantaneous cooling time $\tau_T = -T\left(dT/dt\right)^{-1}$, all of which can be obtained in time-resolved photoemission~\cite{Gierz2013} or multi-color pump-probe experiments~\cite{multi-probe}. Given these quantities at some particular moment of time, it is easy to obtain the true recombination time $\tau_r$ at this moment using the identity
\begin{eq}{cooling-recombination}
    \frac{dn}{dt} &= \frac{\partial n}{\partial \mu}\frac{d\mu}{dt}+\frac{\partial n}{\partial T}\frac{dT}{dt} \\
    \Rightarrow \frac{n - n_{\text{eq}}}{\tau_r} &= \frac{\partial n}{\partial \mu} \frac{\Delta\mu/2}{\tau_{\Delta\mu}}+\frac{\partial n}{\partial T}\frac{T}{\tau_T}
\end{eq}
with $n=n(\mu = \Delta\mu/2, T)$ and $n_{\text{eq}} = n(\mu = 0, T)$. This model can be straightforwardly extended to the case of finite doping. The calculated $\tau_r$ as a function of two dimensionless quantities $\Delta\mu/kT$ and $\tau_T/\tau_{\Delta\mu}$ is shown in \cref{fig:mu-T} in units of $\tau_{\Delta\mu}$ (top row) and $\tau_T$ (bottom row). As can be seen from the top row, $\tau_{\Delta\mu}$ may considerably exceed $\tau_r$, especially in weak nonequilibrium.

\begin{fig}{mu-T}{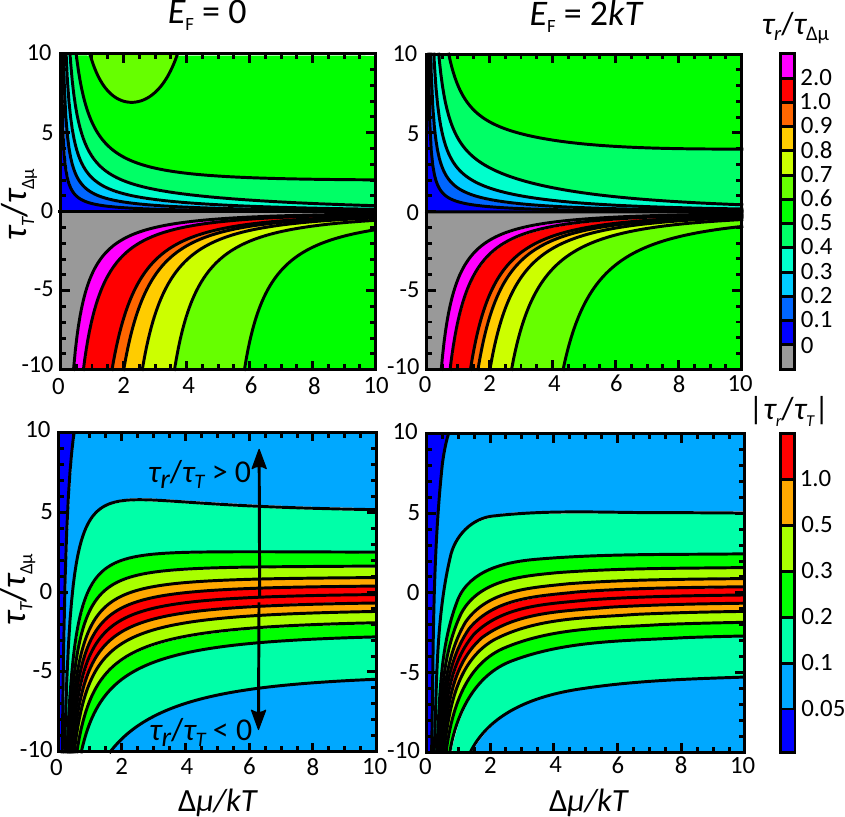}
How to compare theory with experiment: relation between the true recombination time $\tau_r$ in quasiequilbrium graphene, the merging time of quasi-Fermi levels $\tau_{\Delta\mu}$, and the cooling time $\tau_T$. Left column: undoped case; right column: the Fermi energy equals $2kT$, as typically occurs in experiments with photoexcited graphene on SiC.
\end{fig}

There exist some obstacles to an accurate comparison between our theory and time-resolved photoemission experiments. The results presented in this paper do not directly apply to the strongly nonequilibrium case ($\Delta\mu \gtrsim kT$), while small $\Delta\mu$ remain below the experimental energy resolution. However, available experimental data still affords some estimates. At $t \gtrsim 100-200$~fs experiments show no discernible $\Delta \mu$, meaning at least $\Delta \mu \lesssim kT$. Assuming $\tau_{\Delta\mu}$ is positive (Fermi levels converge, otherwise a nonequilibrium distribution would be observed again), we get $\tau_r/\tau_T \lesssim$ 0.2 from the bottom row of \cref{fig:mu-T}. Taking $\tau_T = 700$~fs~\cite{Gierz2013}, we conclude that $\tau_r \lesssim 140$ fs on SiC at $T \sim 1500$~K, in agreement with our calculations, which give $\tau_r = 80$ fs.

\subsection{\label{sec:many-body}Role of many-body phenomena}

\begin{fig}{broadening}{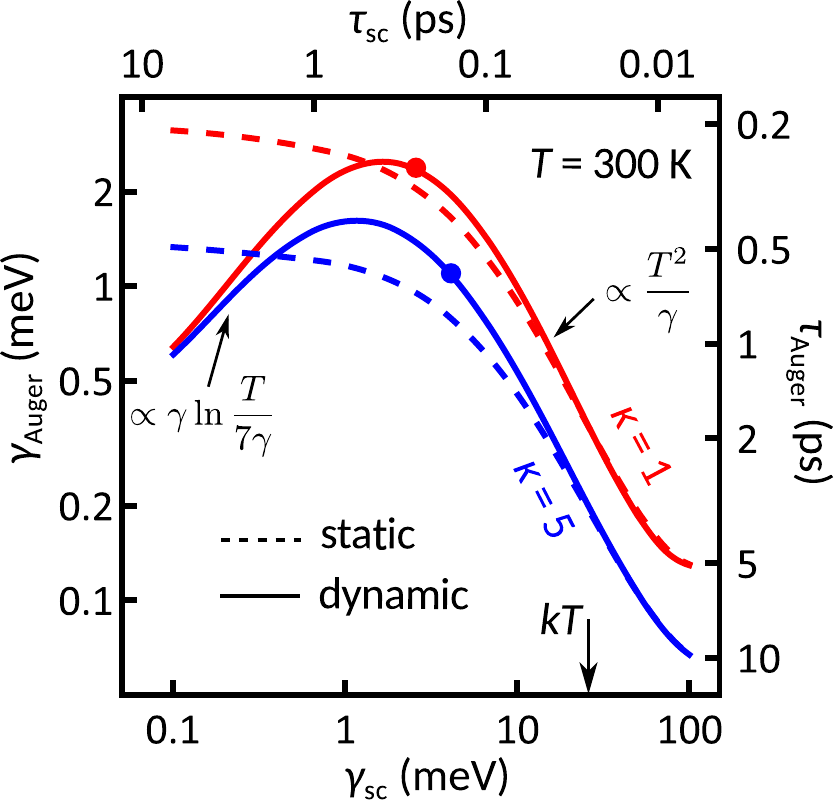}
Impact of the quasiparticle spectrum broadening $\gamma_{\text{sc}}$ (referred to as simply $\gamma$ in the text) on the Auger recombination time in graphene, $\tau_{\text{Auger}}$, at 300 K and two different dielectric constants, $\kappa = 1$ (red lines) and $\kappa = 5$ (blue lines). To make the relation between scattering and recombination clearer, the scattering time $\tau_{\text{sc}} = \hbar\gamma^{-1}_{\text{sc}}$ and the ``Auger broadening'' $\gamma_{\text{Auger}} = \hbar\tau^{-1}_{\text{Auger}}$ are also shown. Dashed (solid) lines: the data calculated with static (dynamic) screening. The results were obtained from \Eqref{ImPccImPcv}; a constant $\Sigma^R_s(\vec{p},E) = -i\gamma_{\text{sc}}$ instead of its $GW$ value was used for calculation of the polarizabilities. Cutoff $\Lambda = 2.5$~eV. Actual $\gamma_{\text{sc}}$ at energy ${\sim} kT$ obtained in our $GW$ calculations are shown with circles.
\end{fig}

Finally, we examine the influence of different many-body phenomena on the AR rate in graphene. The main motivation of this work was to incorporate actual spectrum broadening into the calculations instead of a rough estimate used in \Cite{Tomadin-theory}, as the AR rate is expected to depend crucially on the precise value of the spectrum broadening (when the latter tends to zero, the AR rate vanishes). In \cref{fig:broadening}, we present the AR time in graphene with a constant broadening $\gamma$ included into the spectral function $A_s(p,E) = (\gamma/\pi)/[(E - sv_0p)^2+\gamma^2]$. In the $\gamma \rightarrow 0$ limit, only collinear Auger processes are allowed by conservation laws. When screening is considered in the static approximation, $\varepsilon(q,\omega)\approx\varepsilon(q,0)$ (dashed lines), these processes experience finite screening, and the AR rate has a nonzero value at $\gamma \rightarrow 0$~\cite{Rana-Auger,Tomadin-theory}. In contrast, when the full frequency dependence of the dielectric function is considered (solid lines), the infinite screening of collinear processes makes the AR rate vanish as $\gamma \ln (CkT/\gamma)$ with a constant $C$ of order unity (see Appendix~\ref{sec:broadening}), independently of $\kappa$ (intrinsic screening dominates over extrinsic).

It might seem surprising at first that at larger $\gamma$ the inverse AR time decreases again. However, this behavior is easy to understand by considering the $\gamma \gg kT$ limit, when the density of nonequilibrium carriers in a strongly broadened Dirac cone grows as $n_{\text{noneq}} \propto \gamma \Delta\mu$, while the AR rate remains constant (three frequency integrations with the Fermi-Dirac distributions give $R_{\text{Auger}} \propto T^2\Delta\mu$, see Appendix~\ref{sec:broadening}). In this limit, the characteristic momentum transfer in AR, $\hbar q \sim \gamma$, is much greater than the characteristic energy transfer $\hbar\omega \sim kT$, justifying the static screening approximation (solid and dashed lines merge). Another unusual feature of the large-$\gamma$ case is the enhancement of AR when dynamic screening is used instead of static one. This is attributed to plasmon-enhanced AR in the $(q,\omega)$ regions around the plasmon dispersion where $\left|\varepsilon(q,\omega)\right| < \left|\varepsilon(q,0)\right|$. We have performed $GW$ calculations with $|\varepsilon(q,\omega)|^{-1}$ clipped at its static value (dashed red curve in \cref{fig:methodscomparison}) and found that plasmon-enhanced processes account for around 40\% of the total AR rate (solid red curve in \cref{fig:methodscomparison}). Since plasmon-enhanced AR can be viewed as electron-hole recombination with emission of plasmons followed by their reabsorption by other carriers, it can be argued that letting plasmons decay into free electromagnetic modes instead of getting reabsorbed might open a route towards efficient graphene-based THz lasers. This can be achieved by increasing the coupling between plasmons and free electromagnetic modes (e. g., by plasmonic gratings).

\begin{fig}{methodscomparison}{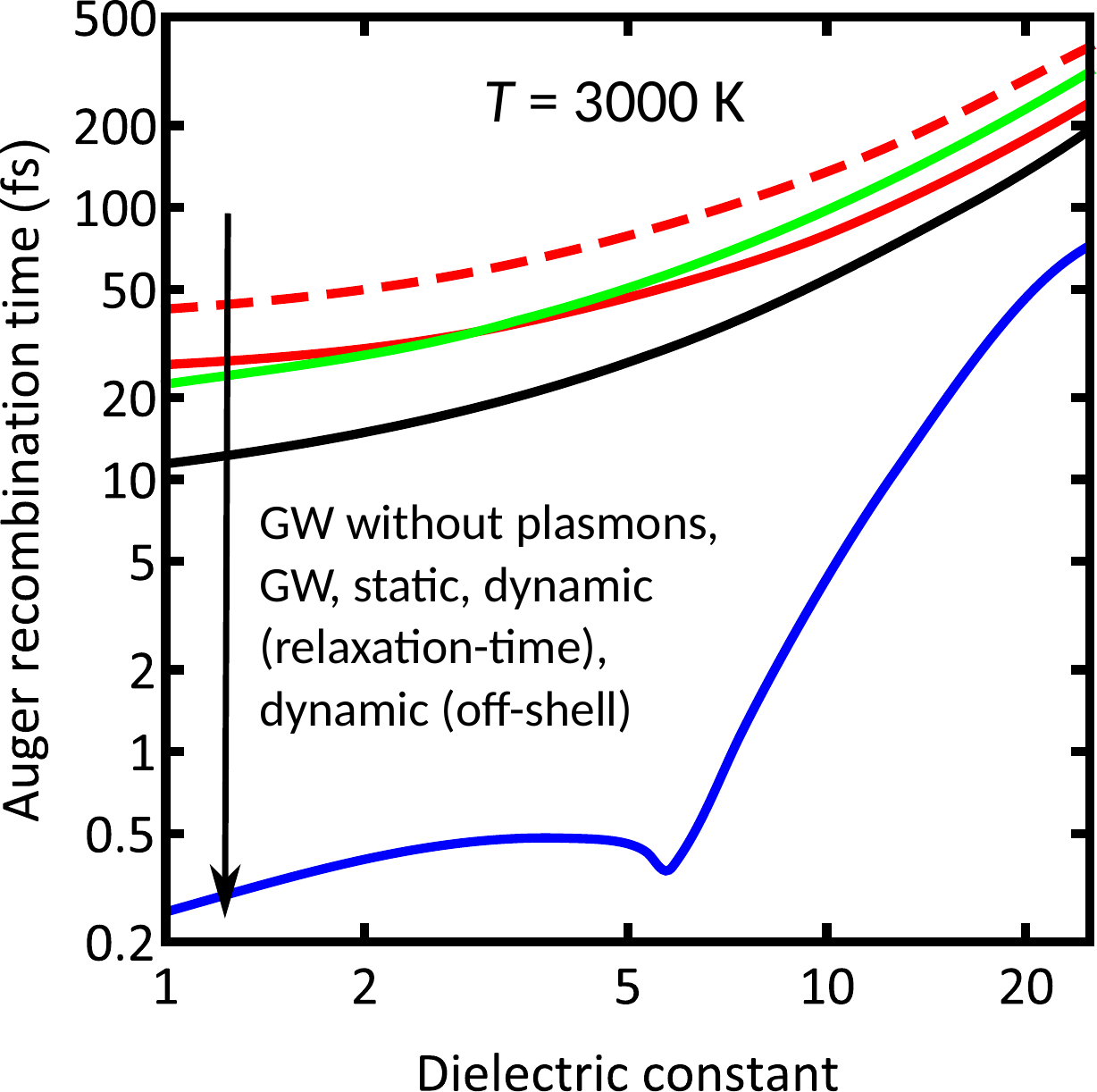}
Impact of different approximations for screening on the Auger recombination time in graphene. Solid red curve: our $GW$ calculations. Dashed red curve: $GW$ calculations with $\varepsilon(q,\omega) \rightarrow \max(\varepsilon(q,\omega),\varepsilon(q,0))$ substitution in \eqref{ImPccImPcv}. Green, blue, black curves were calculated using \Eqref{regularizedrate} for an infinite Dirac cone with infinitesimal broadening~\cite{Rana-Auger,Tomadin-theory} and different models of screening. Green curve: static screening, $\varepsilon(q,q) \rightarrow \varepsilon(q,0)$~\cite{Rana-Auger}. Blue curve: dynamic screening with spectrum broadening $\gamma$ incorporated through ``off-shell regularization'' $|\varepsilon(q,q)|^{-2} \rightarrow (1/2)\left\{|\varepsilon(q,q+2\gamma)|^{-2}+|\varepsilon(q,\max(0,q-2\gamma))|^{-2}\right\}$ (in analogy to \Cite{Tomadin-theory}). Black curve: dynamic screening within relaxation-time approximation $\varepsilon(q,q) \rightarrow \varepsilon(q,q+2i\gamma)$ (in analogy to \Cite{Malic_dynamic}). $\gamma$ was taken from our $GW$ calculations as $-\Im\Sigma^R$ on shell at energy $kT$. High temperature $T = 3000$~K ensures the possible cutoff-dependent artifacts of $GW$ approximation are minimal (cutoff $\Lambda = 2.5$~eV).
\end{fig}

We also demonstrate the unreliability of simplistic ways to include the spectrum broadening and screening into the AR rate calculations. In previous works on AR in graphene, spectrum broadening was usually not considered explicitly (with exception of \Cite{Malic_broadening}). Instead, in \Cite{Rana-Auger} the limit of zero spectrum broadening was considered within the static screening approximation, and \Eqref{regularizedrate} comprising only strictly collinear processes was obtained. Later, the same expression was used with dynamic screening, and spectrum broadening $\gamma$ was taken into account either by taking the dielectric constant for slightly noncollinear processes, assuming some particles are off shell ($\varepsilon(q,q\pm2\gamma)$ instead of $\varepsilon(q,q)$)~\cite{Tomadin-theory}, or by using an advanced version of the relaxation-time approximation ($\varepsilon(q,q+2i\gamma)$ instead of $\varepsilon(q,q)$)~\cite{Malic_dynamic}, where $\gamma$ is not constant and is calculated from in- and out-scattering rates for each electron state.

Green, blue, and black curves in \cref{fig:methodscomparison} show the results obtained using \Eqref{regularizedrate} within different models of screening: static~\cite{Rana-Auger} (green), ``off-shell dynamic''~\cite{Tomadin-theory} (blue), and ``relaxation-time dynamic''~\cite{Malic_dynamic} (black). We use a constant $\gamma$ obtained from our $GW$ calculations as $-\Im\Sigma^R$ on shell at energy $kT$ (the results do not change significantly if a non-constant $\gamma$ is used in the ``relaxation-time dynamic'' approximation, as prescribed in \Cite{Malic_dynamic}). Surprisingly, we find that the static screening approximation gives results rather close to the full $GW$ calculations. This indicates that the enhanced screening of collinear processes and the weakened screening of plasmon-assisted processes approximately average to the static value of screening. The introduction of dynamic screening increases the AR rate despite the collinear divergence of unbroadened polarizabilities: at realistic broadening, the ``dynamical anti-screening'' effects reach the collinear region, at least in the undoped case. The ``off-shell dynamic'' screening approximation shows an overall poor performance, especially at low $\kappa$ (large $\gamma$), because at some wavevectors $\varepsilon(q,q+2\gamma)$ is taken precisely at the plasmon dispersion, and the contribution of such processes is therefore strongly overestimated. It can be cured by considering only $\varepsilon(q,q-2\gamma)$, but at the price of completely neglecting the plasmon contributions. On the other hand, the ``relaxation-time dynamic'' approximation shows a better performance, but still worse than the static one.

The successful application of the static screening approximation to AR in graphene in previous works~\cite{Rana-Auger,Malic-kinetics,Malic-Auger,Malic2017} can also be explained by looking at \cref{fig:broadening}: if actual values of $\gamma$ obtained in self-consistent calculations are used (shown with circles), the results with dynamic screening turn out to be rather close to the results with static screening in the $\gamma \rightarrow 0$ limit. However, other many-body effects render this early approach unreliable in some range of parameters (e.g., at room temperature---compare \cref{fig:broadening} and \cref{fig:substrate}).

\begin{fig}{renormalization}{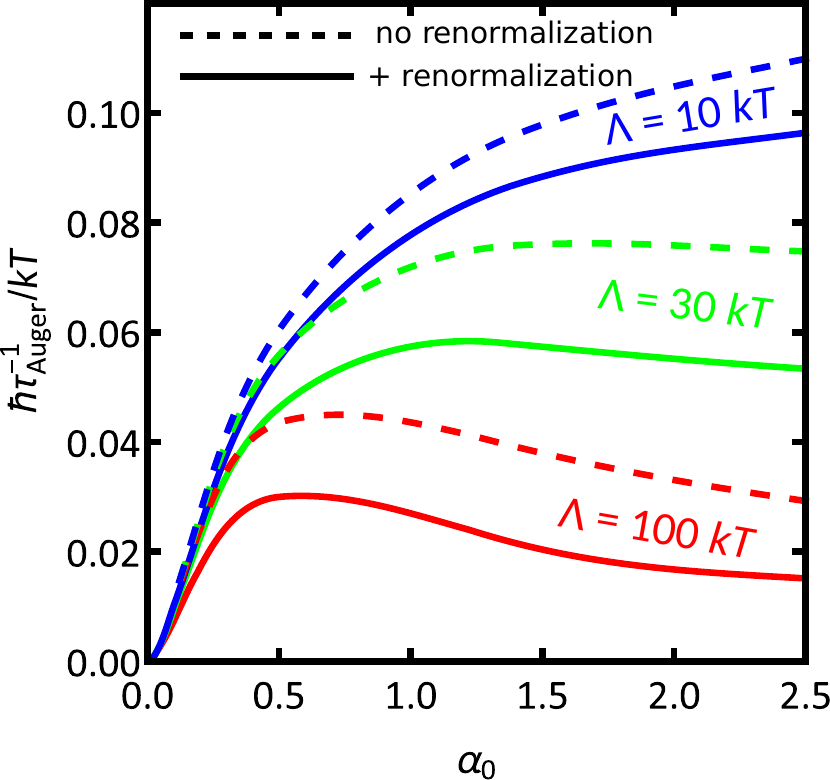}
Solid lines: the same as in Fig.~\ref{fig:dimensionless}; dashed lines: the AR rate without renormalization effects ($\Sigma^R_s(p,E) \rightarrow \Sigma^R_s(p,E) - \Re \Sigma^R_s(p,\epsilon_{s\vec{p}})$ substitution was done at each $GW$ iteration).
\end{fig}

These effects are the renormalization of the quasiparticle spectrum and of the screened interaction. The role of the former is shown in \cref{fig:renormalization}. At small $\kappa$ and $T$ spectrum renormalization can reduce the AR rate by a factor of two. We mention that it affects the AR rate not only by strengthening the conservation law constraints, but also by reducing the coupling constant.

In real materials, nonlinear dispersion does not require electron-electron interactions and is present already in a tight-binding model. The impact of this nonlinearity on the AR time depends on the relative magnitude of this nonlinearity compared to the spectrum broadening. If the nonlinearity is smaller, it will not have much effect on the conservation-law restrictions and, therefore, on the AR time. In graphene, this effect is rather small up to moderately high temperatures: e.g., at $T = 1000$~K the true asymmetric electron-hole dispersion deviates from the Dirac cone by $\sim 1.3$~meV at $E = kT$~\cite{e-h_asymmetry}, which is less than the spectrum broadening $\gamma \sim 13$~meV at $\kappa = 5$. Trigonal warping is even weaker~\cite{Golub-valley} (${\sim} 0.4$~meV at the above parameters), and its influence becomes further reduced after averaging over angles. In contrast, in CdHgTe quantum wells deviations from Dirac dispersion become significant already at several tens of meV~\cite{CdHgTe_bandstructure} and should play a crucial role in AR at room temperature.

The renormalization of the screened interaction leads to the cutoff dependence of the AR rate and produces some unexpected features in the previous plots (e.g., nonmonotonicity in \cref{fig:substrate} and intersecting curves in \cref{fig:gate}). It essentially originates from long tails of the spectral function, which translate into long tails of the imaginary part of the intraband polarizability. This affects the real part of the latter via the Kramers-Kronig relations and therefore enhances screening, suppressing AR. Its influence is most pronounced at large $\Lambda/(kT)$---that is, at low temperatures---and at strong coupling.

The reader is probably surprised by such a strong cutoff dependence of observable quantities. We would like to clarify that this is most likely an artifact of the chosen approximation. Indeed, incorrect high-frequency behavior of $\Im \Pi^R$ violating sum rules is a known drawback of the self-consistent $GW$ approximation~\cite{GW_sum_rules_violation}. We have tried to include spectrum broadening only in $\Im \Pi^R$, but not in $\Re \Pi^R$, thereby eliminating the strong cutoff dependence from the latter (causality was restored in the self-energy by using Kramers-Kronig relations for calculating $\Re \Sigma^R$), and the results indicate that one should probably use low cutoffs (say, $\Lambda = 5 kT$) within the $GW$ scheme to obtain accurate results. Another argument in favor of using a low cutoff is that $\Im \Pi^R$ tails become increasingly longer at high cutoffs, leading to an increasingly stronger violation of sum rules, implying the $GW$ solution moves away from the exact one. However, holding $\Re \Pi^R$ unbroadened is a rather \textit{ad hoc} trick, while an accurate resolution of the cutoff problem would require going beyond the $GW$ approximation and including vertex corrections, which is a computationally challenging task. Therefore, we leave this problem for future studies.

Finally, we comment on the validity of our results in view of the cutoff problem. Most of our qualitative results are cutoff-independent and have a simple physical explanation (those that do not---e.g., the nonmonotonicity in \cref{fig:dimensionless}---are not presented as our main results), therefore they are probably correct. The comparison with experiments is performed at high temperature, where the results are justified by the low $\Lambda/kT$ ratio, as discussed above. On the other hand, this is not the case for our room-temperature AR times, which might therefore prove too long.
\section{\label{sec:conclusions}Conclusions}
We have developed a general formalism for the calculation of Auger recombination rates in Dirac materials, where the electron-hole dispersion is quasi-relativistic. In this case, the lowest-order Auger processes are prohibited by the energy-momentum conservation laws, calling for many-body effects to be considered.  We have shown that within the $GW$ approximation, the AR rate can be expressed as the product of the inter- and intraband polarizabilities timed by the square of the screened Coulomb interaction [\Eqref{ImPccImPcv}], and these \emph{dressed} quantities conveniently absorb the effects of carrier interaction.

Using the developed formulation, we have shown that  AR in graphene is very sensitive to many-body effects, and their accurate treatment is essential to obtain the correct AR rate. These many body effects include: (i) Broadening of the quasiparticle spectrum due to their finite lifetime. As the scattering time $\tau_{\text{sc}}$ becomes long, the AR time scales as  $\tau_{\text{Auger}} \propto \tau_{\text{sc}}/\ln(\text{const}\,\times\,T\tau_{\text{sc}})$ and tends to infinity in the absence of scattering. In practice, the limit of long scattering time is not realized, and the energy-momentum restrictions for AR are considerably washed out. (ii) The interaction-driven velocity enhancement near the Dirac point. It implies the reduction in the effective interaction constant and prolongation of the recombination time. This effect becomes most important at low dielectric constants and temperatures. (iii) Plasmon-assisted recombination. Accurate treatment of dynamic screening shows that it not only suppresses collinear Auger processes, but also increases the contribution of noncollinear processes by anti-screening near the plasmon dispersion, so that the AR rate may turn out even larger than in the static screening model, contrary to the expectations based on the collinear divergence of the bare polarizabilities~\cite{Tomadin-theory}. 

The calculated recombination time $\tau_r$ in undoped graphene is around 1--2 ps at room temperature on most common dielectrics and below 100 fs at ${\sim}$1500--3000 K, which are the typical carrier temperatures in photoexcited graphene. We demonstrate that the experimentally observed lifetime of population inversion $\tau_{\Delta \mu}$ is longer than the recombination time (the lifetime of excess carrier density) $\tau_r$ due to the effect of carrier cooling, which effectively pushes the quasi-Fermi levels apart. Taking this into account, we show that the available experimental data~\cite{Gierz2013} restricts the possible values of the recombination time in graphene on SiC to $\tau_r \lesssim 140$~fs at 1500 K, in agreement with our calculations, which yield $\tau_r = 80$~fs.

Our results show that the AR time in graphene depends on the background dielectric constant $\kappa$ at most linearly within the experimentally relevant range. This limits the possibilities to control the recombination rate in graphene by the choice of substrate, especially if one takes into account that high-$\kappa$ dielectrics provide an additional recombination channel through substrate polar phonons~\cite{Rana-plasmons}. However, AR can be readily suppressed by placing the graphene layer near a metal gate, which increases the room-temparature AR time from 1 ps at $\kappa = 5$ without gate to 7 ps with gate at 1 nm from graphene. An even more substantial AR suppression can be achieved in gapped graphene with $E_g > kT$. The AR time scales according to the Arrhenius law, as $\tau^{-1}_{\text{Auger}} \propto \exp(-E_{\text{th}}/kT)$ with $E_{\text{th}}$ close to $E_g$ or slightly more, yielding $\tau_{\text{Auger}} \approx 30$~ns  at $\kappa = 5$, $T = 300$~K, and $E_g = 0.2$~eV (similar parameters are realized in graphene epitaxially grown on SiC~\cite{Substrate-induced_gap}).

As the screening of Coulomb scattering by metal gates was found to have a strong effect on AR, one might expect the screening by adjacent graphene layers in multilayer stacks also to have a strong effect on the carrier dynamics. However, extra graphene layers not only contribute to screening, but also provide an extra pathway for recombination. Indeed, the energy and momentum released upon electron-hole annihilation in one layer can be transferred to an electron in an adjacent layer. A similar process has been studied in the context of interlayer heat transfer~\cite{Mihnev_InterlayerCooling}, and can be easily included in our formalism. Due to the competition between screening and extra recombination channels, it is \textit{a priori} not clear whether adjacent layers would prolong or shorten the recombination time.

An important future extension of our method would be to go beyond the $GW$ approximation and study the role of vertex corrections, both in the polarizability and the self-energy. The neglect of vertex corrections to the self-energy in \Eqref{Kadanoff-Baym} amounts to the neglect of exchange terms in the AR rate, which is justified when the number of fermion flavors $g$ is large, as is the case in certain Dirac materials. Even in graphene, with $g=4$, the neglect of exchange actually produces much smaller errors than $1/g = 25$~\%~\cite{Tomadin-theory}, while in the recently discovered three-dimensional Weyl semimetal TaAs $g$ is as large as 16~\cite{Xu_Science_TaAs}. However, in three-dimensional topological insulators typically $g=1$~\cite{topinsulatorsreview}, and the exchange terms are expected to play a significant role. On the other hand, vertex corrections to the polarizabilities are generally of the same order as the corrections resulting from ``dressing'' the Green's functions and partially cancel them~\cite{cancellation}, which might cure the unexpected strong-coupling behavior of the AR rate arising from the cutoff dependence of the intraband polarizability.

Another important extension is the study of recombination at strong pumping, which is often realized in photoexcited graphene and injection lasers. This extension is not straightforward. The reason is that a quasi-equilibrium carrier distribution with a high enough population inversion $\mu_c - \mu_v$ may act as a gain medium for certain plasmon modes, which is unphysical in steady-state case and renders \Eqref{ImPccImPcv} divergent ($\epsilon^{R}(q,\omega)$ turns to zero at some points). This divergence of Coulomb scattering was noted already in the kinetic theory of unstable plasma~\cite{RogisterUnstable} and led to an overestimate of scattering rates in the early simulations of pump-probe experiments~\cite{Scott_PRL_PlasmonUndamping}.  
Thus, the case of a continuous strong pump requires a full pump$+$resonator$+$active medium simulation~\cite{Koch_PRA_ManyBodyLaser}, so that the plasmonic gain is either removed by spectral hole burning in the carrier distribution or balanced by loss due to plasmons leaving the active region, and the divergence disappears. This divergence also disappears in the case of a pulsed pump, but time-dependent simulations of carrier kinetics with taking into account quantum many-body effects (which are essential to describe AR) require the use of two-time Green's functions~\cite{Haug} and are computationally demanding.
\begin{acknowledgments}
This work was supported by the grant No. 16-19-10557 of the Russian Science Foundation. The authors are grateful to T. Otsuji, V. Aleshkin and M. Bonitz for helpful discussions.
\end{acknowledgments}
\newpage
\appendix

\section{\label{sec:broadening}Dependence of the AR rate in graphene on the spectrum broadening}

At $\gamma \ll kT$, the main contribution to the AR rate is from the $\omega \approx q$ region, where the polarizabilities exhibit square-root singularities~\cite{Finite-temperature_polarizability}, $\Pi^{R}_{cc/vv} \propto 1/\sqrt{\omega^2-q^2}$ and $\Pi^{R}_{vc} \propto 1/\sqrt{q^2-\omega^2}$ ($\Pi^{R}_{cv}$ is smooth there), while the screened Coulomb interaction $W^R \approx -1/\Pi^R$ approaches zero. After factoring out the square roots, the remaining integrand in \Eqref{ImPccImPcv} stays finite (divergences of the Bose functions are cancelled by zeros of the polarizabilities) and gives a $\gamma$-independent factor. $cc \rightarrow vv$ process is strongly suppressed by conservation laws and can be neglected. $\omega \rightarrow \omega + i\gamma$ substitution in the retarded Green's function approximately translates into $\omega \rightarrow \omega + 2i\gamma$ substitution in the polarizabilities, and \Eqref{ImPccImPcv} becomes
\begin{eq}{lowgammarate}
    R_{\text{Auger}} &\propto \sum_{\vec{q},\omega} \Im \frac{1}{\sqrt{(\omega+2i\gamma)^2-q^2}} \Im \frac{1}{\sqrt{q^2 - (\omega+2i\gamma)^2}}\\
    &\times \left| (\omega+2i\gamma)^2-q^2 \right| \propto \sum_{\vec{q}} \int d \omega \Im \frac{ (\omega+2i\gamma)^2-q^2}{\left| (\omega+2i\gamma)^2-q^2 \right|}\\
    &\propto \sum_{\vec{q}} \int d \omega \frac{\gamma \omega}{\sqrt{(\omega^2-4\gamma^2-q^2)^2+16\omega^2\gamma^2}}\\
    &\propto \sum_{\vec{q}} \gamma \left. \arcsinh \frac{\omega^2+4\gamma^2-q^2}{4\gamma q} \right|_{\omega_{\text{min}}}^{\omega_{\text{max}}}.
\end{eq}

The $\omega$ and $\vec{q}$ integrals are cut off at $\sim \pm kT$ by the Bose functions and the Coulomb interaction (the latter decays like $1/q$ at $q \gg \omega, kT$), and \eqref{lowgammarate} yields $R_{\text{Auger}} \propto \gamma \ln (CkT/\gamma)$ with $C$ of order unity (numerical calculations depicted in Fig.~\ref{fig:broadening} yield $C \approx 0.14$).

To provide an estimate of $R_{\text{Auger}}$ in the $\gamma \gg kT$ case, it is more convenient to rewrite \Eqref{ImPccImPcv} in terms of spectral functions~\cite{Yevick-GW_Auger} in a form resembling the Fermi Golden Rule:
\begin{eq}{highgammarate}
     R_{\text{Auger}} &= 2\pi g^2 \sum_{\vec{p}_{1,2,3,4}} \int dE_{1,2,3,4}\\
    &\left| W^R(\vec{p}_1-\vec{p}_3,E_1-E_3) \right|^2 u^{cv}_{\vec{p}_1,\vec{p}_3}u^{cc}_{\vec{p}_2,\vec{p}_4}\\
    &\times A_c(\vec{p}_1,E_1)A_c(\vec{p}_2,E_2)A_v(\vec{p}_3,E_3)A_c(\vec{p}_4,E_4)\\
    &\times \left\{ f_c(E_1)f_c(E_2)[1-f_v(E_3)][1-f_c(E_4)] \right.\\
    &- \left. [1-f_c(E_1)][1-f_c(E_2)]f_v(E_3)f_c(E_4) \right\}\\
    &\times (2\pi)^2 \delta \left(\vec{p}_1+\vec{p}_2-\vec{p}_3-\vec{p}_4\right)\\
    &\times \delta \left(E_1+E_2-E_3-E_4\right).
\end{eq}

Now we examine it term by term. The delta functions remove one momentum and one energy integration. The distribution functions constrain the remaining frequency integrals to $|E_i|\lesssim kT$. The momentum integrals are constrained to $p_i\lesssim \gamma$ by the spectral functions. The screened Coulomb interaction can be taken in the static approximation since $p_i \sim \gamma \gg E_i \sim kT$, and is of order $1/\gamma$. Finally, the overlap factors lie between 0 and 1 and are of order 1 on average. In view of the above, at $\gamma \gg kT$ the pure recombination term is proportional to $\gamma^6 T^3 \gamma^{-2} \gamma^{-4} = T^3$. After subtracting generation, we get $R_{\text{Auger}} \propto T^2 (\mu_c-\mu_v)$. On the other hand, the density of nonequilibrium carriers acquires a $\gamma$ dependence: $n_{\text{noneq}} \propto \iint d^2\vec{p} \int dE A(\vec{p},E)[f_c(E)-f_c^{eq}(E)] \propto \gamma^2 T \gamma^{-1} (\mu_c-\mu_v)/T = \gamma(\mu_c-\mu_v)$, so the inverse recombination time scales as $T^2/\gamma$.

In the above arguments all the integrals were assumed to be convergent. Actually, this is not the case, because the density of states of a 2D Dirac cone with constant broadening diverges logarithmically: $\text{DOS}(E=0)=g\iint \frac{d^2\vec{p}}{(2\pi)^2} A(p,E=0)\approx \frac{g}{2\pi^2} \gamma \ln \frac{\Lambda}{\gamma}$, adding a cutoff dependence to the above estimates. However, it does not change the main trends, and the estimates derived in this section comply with numerical calculations.
\section{\label{sec:dielectric}Model dielectric functions}

The phonon-induced frequency dependence of a dielectric function is usually described by the Lorentz oscillator model:
\begin{eq}{Lorentz}
    \kappa(\omega) = \kappa_{\infty} + \sum_{i=1}^{N} \frac{(\kappa_{i-1}-\kappa_i)\omega^2_i}{\omega^2_i-\omega(\omega + i\gamma_i)},
\end{eq}
where $\omega_i$ and $\gamma_i$ are the frequencies and the damping constants of $N$ transverse optical modes, and $\kappa_i$ are the intermediate dielectric constants related to the oscillator strengths and the frequencies of longitudinal modes ($\kappa_0$ is the static dielectric constant; $\kappa_N \equiv \kappa_{\infty}$ is the high-frequency dielectric constant). The parameters we used are presented in \cref{tab:phonons}.

\begin{table}[!b]
\caption{\label{tab:phonons}Parameters of the dielectrics considered in this article.}
\begin{ruledtabular}
\begin{tabular}{dddd}
\multicolumn{1}{c}{$i$} & \multicolumn{1}{c}{$\omega_i$ (meV)} & \multicolumn{1}{c}{$\gamma_i$ (meV)} & \multicolumn{1}{c}{$\kappa_i$}\\ \hline
\multicolumn{4}{c}{HfO$_2$\footnote{\CCite{HfO2kappa}.}}\\
0 & & & 14.2\\
1 & 23.2 & 26.8 & 12.4\\
2 & 31.6 & 5.6 & 10.3\\
3 & 41.8 & 7.7 & 6.2\\
4 & 50.0 & 7.0 & 4.4\\
5 & 62.7 & 6.7 & 3.9\\
6 & 73.8 & 3.2 & 3.8\\ \hline
\multicolumn{4}{c}{hBN\,\footnote{\CCite{hBNkappa}.}}\\
0 & & & 7.0\\
1 & 95.1 & 4.3 & 6.8\\
2 & 169.5 & 3.6 & 5.0\\ \hline
\multicolumn{4}{c}{6H-SiC\,\footnote{\CCite{SiCkappa}. Note that we consider graphene \emph{on} SiC and take $\kappa(\omega)=(\kappa_{\text{SiC}}(\omega)+\kappa_{\text{air}}(\omega))/2=(\kappa_{\text{SiC}}(\omega)+1)/2$. The dielectric constants listed in this Table apply to $\kappa_{\text{SiC}}(\omega)$.}}\\
0 & & & 10.0\\
1 & 98.4 & 0.6 & 6.7\\
\end{tabular}
\end{ruledtabular}
\end{table}


%\bibliography{Bibliography}

\end{document}